\documentclass{aa}

\usepackage{natbib}
\bibpunct{(}{)}{;}{a}{}{,} 
\usepackage{graphicx}   
\usepackage[varg]{txfonts}
\usepackage{hyperref}   

\hypersetup{
    colorlinks=true,       
    linkcolor=blue,          
    citecolor=blue,        
    filecolor=blue,      
    urlcolor=blue           
}

\def\ga{\,\,\raise0.14em\hbox{$>$}\kern-0.76em\lower0.28em\hbox{$\sim$}\,\,}
\def\la{\,\,\raise0.14em\hbox{$<$}\kern-0.76em\lower0.28em\hbox{$\sim$}\,\,}

\def\iso#1{$^{#1}$}
\def\Msun{$M_{\odot}$}

\def\msun{M_{\odot}}
\def\pg{$p,\gamma$}

\def\cm3{cm$^{-3}$}
\def\an{$\alpha,n$}

\def\chem#1#2{$\mathrm{^{#2}\kern-0.8pt#1}$}
\def\reac#1#2#3#4#5#6{$\mathrm{\, ^{#2}\kern-0.8pt{#1}\, ({#3}\, ,{#4})\, {}^{#6}\kern-0.8pt{#5}\, }$}

\def\conv{\mathrm{conv}}

\def\nuc{\mathrm{nuc}}

\def\be{\begin{equation}}
\def\ee{\end{equation}}
\def\beqy{\begin{eqnarray}}
\def\eeqy{\end{eqnarray}}
\def\bmlet{\begin{mathletters}}
\def\emlet{\end{mathletters}}

\begin{document}

\title{The intermediate neutron capture process}
\subtitle{II. Nuclear uncertainties}

\author{S. Goriely
\and L. Siess
\and A. Choplin}
\offprints{sgoriely@astro.ulb.ac.be}

\institute{
Institut d'Astronomie et d'Astrophysique, Universit\'e Libre de Bruxelles,  CP 226, B-1050 Brussels, Belgium
}

\date{Received --; accepted --}

\abstract
{
Carbon-enhanced metal-poor (CEMP) r/s-stars show surface-abundance distributions characteristic of the so-called intermediate neutron capture process (i-process) of nucleosynthesis. 
We previously showed that the ingestion of protons in the convective helium-burning region of a low-mass low-metallicity star can explain the surface abundance distribution observed in  CEMP r/s stars relatively well. Such an i-process requires detailed reaction network calculations involving hundreds of nuclei for which reaction rates have not yet been determined experimentally.
}
{
We investigate the nuclear physics uncertainties affecting the  i-process during the asymptotic giant branch (AGB) phase of low-metallicity low-mass stars by propagating the theoretical uncertainties in the radiative neutron capture cross sections, as well as the  \iso{13}C(\an)\iso{16}O reaction rate, and estimating their impact on the surface-abundance distribution.
}
{
We used the {\sf STAREVOL} code to follow the evolution of a 1 \Msun\ [Fe/H]~$=-2.5$  model star during the proton ingestion event occurring at the beginning of the AGB phase. In the computation, we adopt a nuclear network of 1160 species coupled to the transport processes and different sets of radiative neutron capture cross sections consistently calculated with the  {\sf TALYS} reaction code. 
}
{
It is found that considering  systematic uncertainties on the various nuclear ingredients affecting the radiative neutron capture rates, surface elemental abundances are typically predicted within $\pm 0.4$~dex. The  $56 \la Z \la 59$ region of the spectroscopically relevant heavy-s elements of Ba-La-Ce-Pr as well as the r-dominated Eu element remain relatively unaffected by nuclear uncertainties. In contrast, the inclusion of the direct capture contribution impacts the rates in the neutron-rich $A\simeq 45$, 100, 160, and 200 regions, and the i-process production of the $Z\simeq 45$ and 65-70 elements. Uncertainties in the photon strength function also impact the overabundance factors by typically 0.2-0.4 dex. Nuclear level densities tend to affect abundance predictions mainly in the $Z=74-79$ regions.
The uncertainties associated with the neutron-producing reaction \iso{13}C(\an)\iso{16}O and the unknown $\beta$-decay rates are found to have a low impact on the overall surface enrichment.
}
{
The i-process nucleosynthesis during the early AGB phase of low-metallicity low-mass stars remains sensitive to nuclear uncertainties, substantially affecting theoretical predictions of still unknown radiative neutron capture cross sections.  Improved descriptions of direct neutron capture based on shell model calculations or experimental constraints from ($d,p$) reactions could help to decrease the uncertainties in the estimated rates. Similarly, constraints on the photon strength functions and nuclear level densities, for example through the Oslo method, in the neutron-rich region of $A\simeq 100$ and 160 would increase the predictive power of the present simulations.
}

\keywords{nuclear reactions, nucleosynthesis, abundances -- stars: AGB and post-AGB}

\titlerunning{Development of the i-process in low-metallicity AGB stars}

\authorrunning{Authors }

\maketitle


\section{Introduction}
\label{sect:intro}
Elements heavier than iron are mainly produced by neutron capture reactions \citep[e.g.][]{Arnould20}.
Two main neutron capture processes are known to be responsible for the production  of almost
all of the trans-iron elements in the Universe; these are the slow and rapid neutron capture processes (also referred to as the s- and r-processes, respectively).
In addition, an intermediate neutron capture process, or i-process, with neutron concentrations intermediate between the s- and r-processes ($N_n \approx 10^{13}-10^{15}$\cm3) has already been proposed by \citet{Cowan77}, but was only recently revived
to explain the surface abundances determined in the so-called carbon-rich metal-poor (CEMP) r/s-stars, whose abundances are difficult to reproduce with solely an s-process, an r-process, or a combination of both \citep[e.g.][]{Jonsell06, Lugaro12,Dardelet14,Roederer16}. The neutron flux needed to explain the i-process nucleosynthesis is expected to originate from protons being ingested in a convective region powered by helium burning. In this event, the protons are entrained by the convective flow to regions of higher temperature ($T > 10^8$K) where they are depleted via the reaction \iso{12}C(\pg)\iso{13}N. The subsequent beta decay of \iso{13}N to \iso{13}C is followed by the reaction \iso{13}C(\an)\iso{16}O which produces high neutron densities up to $N_n \approx 10^{15}$\cm3 \citep[e.g.][]{Siess02,Campbell08,Herwig11}.
Various astrophysical sites have been suggested to be responsible for the i-process nucleosynthesis. These include: the early asymptotic giant branch (AGB) phase of metal-poor low-mass stars, in which the entropy barrier between the hydrogen- and helium-rich zones can be surmounted by the energy released by the thermal pulse \citep{Fujimoto00,Chieffi01,Siess02,Iwamoto04,Cristallo09b,Suda10,Cristallo16,Choplin21}; the core helium flash of very low-metallicity low-mass stars \citep{Fujimoto00,Schlattl01,Suda10,Campbell10, Cruz13}; the very late thermal pulses of post-AGB stars \citep{Herwig11};
rapidly accreting carbon--oxygen (C-O) or oxygen--neon (O-Ne) white dwarfs (RAWDs) in close binary systems \citep{Denissenkov17,Denissenkov19}; and super-AGB stars (7~$\msun \lesssim M_{\rm ini} \lesssim$~10~$\msun$) at low metallicity ($Z\la 10^{-3}$)  \citep{Jones16}.; the so-called dredge-out in super-AGB stars, when, at the end of carbon-burning, a helium-driven convective shell merges with the descending convective envelope \citep{Siess07};
or the helium shell of very low- or zero-metallicity massive stars ($M_{\rm ini} > $~10~$\msun$) during central carbon-burning (or later stages) \citep{Banerjee18, Clarkson18, Clarkson20}.

The likelihood of each of these various scenarios explaining the formation of CEMP-r/s stars was studied by \citet{Abate16} and compared with observations. None of the existing scenarios are free from difficulties in explaining the observed frequency of CEMP-r/s stars.
The contribution of the i-process to the Galactic enrichment remains unclear, though it has been estimated by \citet{Denissenkov17} that the production of Ge-Mo elements could originate from the i-process in RAWDs with an efficiency similar to that of the s-process in low-mass AGB stars.

In our first paper \citep{Choplin21},  we investigated how the i-process develops in a low-metallicity low-mass stellar model, its dependence on the numerics, and the nuclear and chemical signatures of the i-process, and compared our model predictions with abundances in r/s stars.
The specificity of this work is to consider the nucleosynthesis taking place in the same low-mass low-metallicity AGB model as in \citet{Choplin21}, namely a 1~\Msun{}, [Fe/H]~$=-2.5$ model star, and to study the sensitivity of the predictions to the various theoretical nuclear uncertainties. We note that i-process simulations require a large amount of nuclear data, many of which can only be estimated based on theory. The predictive power of nucleosynthesis must therefore be questioned and the impact of nuclear uncertainties on abundance predictions analysed in detail. This is the subject of the present paper.

In Sect.~\ref{sect_mod} we briefly describe the input physics with a specific emphasis on the nuclear reaction network used and the way theoretical uncertainties are treated. In light of the key role of the \iso{13}C(\an)\iso{16}O reaction as the neutron source, Sect.~\ref{sect_13c} studies the impact of this reaction on the i-process nucleosynthesis in our fiducial 1~$\msun$ [Fe/H]~$=-2.5$ model star. In Sect.~\ref{sect_ng} we present a detailed analysis of the model uncertainties still affecting the predictions of experimentally unknown reaction rates and their impact on the production of heavy elements by the i-process. The results are further discussed in Sect.~\ref{sect_disc} to highlight the nuclear priorities in terms of nucleosynthesis predictions. Conclusions are given in Sect.~\ref{sect_conc}.


\section{Input physics and some numerical aspects}
\label{sect_mod}

The present calculations are based on the stellar evolution code  {\sf STAREVOL},  the description of which can be found in \cite{Siess00}, \cite{Siess06}, \cite{Goriely18c}, \cite{Choplin21} and references therein. In the present computations, the model is similar to the one described in \citet{Choplin21}.
In particular, it was shown that high temporal and spatial resolutions
are needed for a proper description of the ingestion mechanism and therefore
of the resulting nucleosynthesis. The time-steps as well as the size of the grid mesh needed to ensure that the integration variables remain correctly discretised must be accurately determined. As discussed in \cite{Choplin21}, the temporal and spatial resolutions can be tuned by two parameters in the  {\sf STAREVOL} code. For the present study, these are set to $\alpha=0.01$ and $\epsilon_{\rm max}=0.04$, respectively. See \citet{Choplin21} for more details and in particular for the definition of these parameters.
Some points of specific relevance to the nucleosynthesis calculations are detailed below.

\subsection{Chemical transport}
\label{sect_over}

During proton ingestion, the mixing timescale becomes comparable to the nuclear timescale associated with the energetic reaction \iso{12}C(\pg)\iso{13}N. In these circumstances, the nucleosynthesis,
in particular related to the neutron capture following the \iso{13}N($\beta^+$)\iso{13}C(\an)\iso{16}O reactions,
 cannot be computed independently of the chemical transport. For this reason, we consider the diffusive mixing scheme  described in \cite{Goriely18c} where a diffusion equation is used to simulate the transport of chemicals
\begin{equation}
\frac{\partial X_i}{\partial t}= \frac{\partial}{\partial m_r}  \left[\,(4 \pi r^2 \rho)^2\,D\ \frac{\partial X_i}{\partial m_r}\right] + \frac{\partial X_i}{\partial t}\bigg|_\nuc \ ,
\label{eq_dif}
\end{equation}
where $D = D_{\conv}+D_{\rm mix}$ is the sum of the diffusion coefficients associated with convection and other mixing mechanisms (such as overshoot), respectively. The last term in Eq.~(\ref{eq_dif}) accounts for the change in composition due to nuclear burning.
In our study, from the onset of the AGB phase onward, the nucleosynthesis and diffusion equations are solved simultaneously once the structure has converged and no additional mixing mechanism is included ($D_{\rm mix}=0$).

Despite the above-mentioned coupling between the chemical transport
and the energetics, it should be emphasised that the present
one-dimensional simulation has limitations and the treatment of the complex proton ingestion and stellar
structure response should ideally be studied through multi-dimensional
simulations, as investigated in similar types of stars  by
\citet{Mocak10,Mocak11,Stancliffe11,Herwig14,Woodward15}. These 3D simulations reveal
that turbulent entrainment at the He convective boundary can reach the
H-rich layer and advect protons downward in high-temperature regions. The subsequent evolution is however still
uncertain because these simulations only cover a short period of time
(a few tens of years) and are limited in terms of spatial resolution.
This may explain for example why \cite{Stancliffe11} did not find a splitting of
the He-driven flash. Acknowledging the uncertainties affecting the evolution of
the stellar structure during a proton ingestion event within the framework of a one-dimensional
simulation, our analysis of the nuclear uncertainties remains nevertheless robust.

\subsection{Nuclear reaction network}
\label{sect_nucnet}
In this paper we consider the network of 1160 species introduced in \citet{Choplin21} that extends to neutron-rich nuclei up to Cf and includes some  2123 nuclear ($n$-, $p$-, $\alpha$-captures as well as $\alpha$-decays), weak (electron captures, $\beta$-decays), and electromagnetic interactions. This network involves all species with a half-life of greater than typically 1s and can consequently describe neutron capture processes characterised by neutron densities of up to approximately $10^{17}$\cm3. Maximum neutron densities of typically $10^{15}$\cm3 are found during the i-process modelled here.

Nuclear reaction rates were taken from the Nuclear Astrophysics Library of the Free University of Brussels\footnote{available at http://www-astro.ulb.ac.be/Bruslib} \citep{arnould06}, and include the latest experimental and theoretical cross sections through the interface tool NETGEN \citep{Xu13}. In particular, all the charged-particle-induced reaction rates of relevance in the hydrogen- and
helium-burning calculations were taken from the NACRE and NACRE-II evaluations \citep{Angulo99,Xu13b} and the STARLIB library \citep{Sallaska13}.
 We note that at low temperature, the non-thermalisation of the isomeric state of \iso{26}Al,\iso{85}Kr,\iso{115}In,\iso{176}Lu,  and \iso{180}Ta is introduced explicitly in the reaction network \citep{Kappeler89,Nemeth94}.
The $(n,\alpha)$ reactions and $\alpha$-decays are also included when relevant, in particular between Bi and Cf isotopes.

When not available experimentally, the Maxwellian-averaged cross sections are calculated with the {\sf TALYS} reaction code \citep{Koning12,Goriely08a}. These concern in particular some 793 radiative neutron captures for isotopes with $14 \le Z \le 93$ which may be affected by significant uncertainties, as discussed in Sect.~\ref{sect_ng}. Of these 793 reactions, 663 are below Po ($Z\le 83$). The {\sf TALYS} calculations are also used systematically to deduce the stellar rates   from the laboratory neutron capture cross sections by
allowing for the possible thermalisation of low-lying states in the target nuclei. Similarly, most of the nuclei included in our reaction network have known $\beta$-decay half-lives \citep{Audi17a}, except some 110 for which theoretical predictions are needed. Tables obtained either from the Gross Theory \citep{Tachibana90} or the relativistic mean-field model plus random phase approximation \citep{Marketin16} are considered in this case. The temperature- and density-dependent $\beta$-decay and electron capture rates in stellar conditions are taken from \citet{Takahashi87} with the update of  \citet{Goriely99}. However, we note that the 106 nuclei concerned by such a temperature and density dependence mainly correspond to nuclei close to the bottom of the valley of $\beta$-stability, which are of special interest to the s-process nucleosynthesis. Such effects for the more neutron-rich nuclei of relevance to the i-process still remain to be investigated.

In contrast to the s-process, the i-process nucleosynthesis involves many more unstable nuclei for which experimental reactions are not available at the present time. As mentioned above, nucleosynthesis calculations resort to theoretical predictions which may still be affected by non-negligible uncertainties. To estimate the impact of such uncertainties on the nucleosynthesis, most of the studies up to now have used Monte Carlo-type simulations \cite[e.g.][]{McKay19}. In this approach, reaction or decay rates are classically modified in a given range, independently of the changes of other reactions, and this modified set of reactions is used to compute the nucleosynthesis. This approach is only appropriate if rates are uncorrelated; this may be the case in particular for experimentally determined cross sections or temperature-dependent $\beta$-decay rates for which the properties are intrinsic to the specific case and do not affect reactions on neighbouring nuclei. However, in the case of theoretically derived cross sections, all rates are predicted by a given model that defines the correlations between all reactions, either locally (e.g. by modifying a given structure property of a specific nucleus) or globally (e.g. by changing the model adopted to describe the structure properties of interacting nuclei or the interaction with nucleons or photons). More specifically, changing a given theoretical rate by a given factor without affecting other rates, either locally or globally, is hard to justify. For example, if a rate is uncertain because of the still unknown mass of the target nucleus, it will affect its production as well as destruction rates through a modification of the corresponding $Q$-value. Consequently, none of the (destruction or production) rates can be changed independently from the others. Similarly, theoretical rates are correlated by the underlying reaction model and the many nuclear models adopted to describe the ingredients of the reaction model, such as masses, nuclear level densities, photon strength function, optical potential, and so on.

It should also be mentioned that the impact of a reaction may be significantly different when studied in a one-zone model or in a `realistic' model composed of a large number of zones and bound to mixing processes between layers that may encounter rather different thermodynamic conditions. In the AGB star model considered here, the i-process nucleosynthesis is followed in typically hundreds of neighbouring layers, each of them characterised by different temperatures and neutron density profiles. Even in the convective regions, the conditions cannot be mimicked by a simple one-zone model. The resulting sensitivity of the nucleosynthesis calculations with respect to nuclear ingredients can consequently reach very different conclusions. In particular, in their one-zone simulations, \citet{McKay19} concluded that the $^{66}$Ni($n,\gamma)$ reaction behaved as a major bottleneck. We checked that, in the present AGB
multi-zone calculation where the neutron density $N_n$ approaches $10^{15}~\rm{cm}^{-3}$,
a modification  of this reaction by a factor of a hundred, up or down,
does not affect the nucleosynthesis  flow, and therefore the
abundance predictions,  in any way. This shows that even if $^{66}$Ni($n,\gamma)$ may be relevant in some specific layers, globally it does not affect the final nucleosynthesis in our specific multi-zone simulation.

Predictions are affected by systematic as well as statistical uncertainties. The former, also referred to as model uncertainties, are known to dominate for unstable nuclei \citep[e.g.][]{Goriely14a}, because there is usually no or very little experimental information with which to constrain the model on those nuclei. Whenever experimental information is available, either for rates or reaction model ingredients, it is considered in the theoretical modelling to constrain the possible range of variation of the model parameters, hence the impact of model as well as parameter uncertainties. However, for exotic nuclei, models of different types, ranging between macroscopic and microscopic approaches \citep[e.g.][]{Arnould07,Goriely15b,Hilaire16, Goriely17a}, may provide rather different predictions.
For this reason, in the present study, global systematic uncertainties will be considered, as described in Sect.~\ref{sect_ng} for the radiative neutron captures. Those global systematic uncertainties are propagated to the calculation of reaction rates, and are consistently applied to the nucleosynthesis simulations to estimate their impact on the i-process yields in our  1~\Msun{}, [Fe/H]=-2.5 model star.

\section{Sensitivity to the  \iso{13}C(\an)\iso{16}O reaction rates}
\label{sect_13c}

\begin{figure}
\includegraphics[width=\columnwidth]{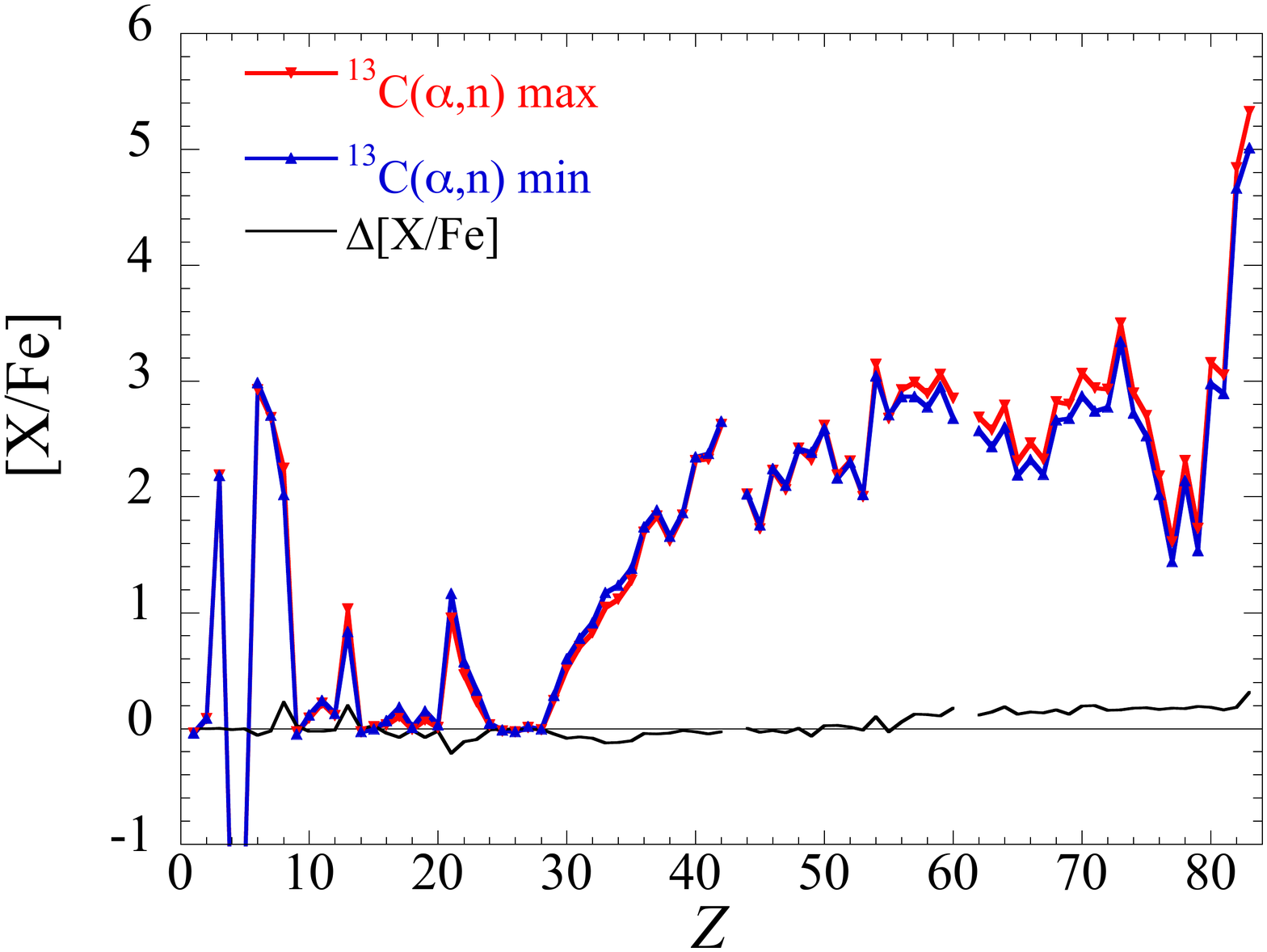}
\caption{Surface elemental overabundances resulting from the i-process in the 1~\Msun{}, [Fe/H]=-2.5 model star when adopting NACRE II upper and lower limits of the \iso{13}C(\an)\iso{16}O reaction rate. The black line corresponds to the difference in the overabundance between the maximum and minimum rate predictions.
 }
\label{fig_ipro_13c}
\end{figure}

The  \iso{13}C(\an)\iso{16}O reaction plays a fundamental role in the production of the neutrons for the i-process. In addition, to enrich the AGB surface with the i-process residuals, this reaction needs to take place before the pulse-driven convective zone splits \citep[see Sect. 3 of ][for more details]{Choplin21}. A change in the  \iso{13}C(\an)\iso{16}O rate may delay or speed up the production of the heavy elements at the bottom of the thermal pulse, thereby affecting the surface enrichment. For this reason we tested its sensitivity, adopting the upper and lower limits of the  \iso{13}C(\an)\iso{16}O reaction rate proposed by the NACRE II evaluation \citep{Xu13b}. The corresponding rate uncertainty is estimated at $\pm 27$\% at the temperature of $T=2.5~10^8$~K, which is characteristic of the i-process at the bottom of the pulse.
Recently, a new analysis of the uncertainties associated with the extrapolation of the cross section at low energies \citep{deBoer20} estimated an even lower uncertainty of  $\pm 16$\% at this temperature.
Figure~\ref{fig_ipro_13c} illustrates the impact of such uncertainties on the surface overabundance in our 1~\Msun{}, [Fe/H]=-2.5 model star. Increasing the experimental value from the lower to the upper limit tends to increase the overall $Z>55$ surface overabundances by a maximum of 0.2 dex.
We tested a larger modification of the NACRE II rate by a factor of two, up and down. In this case, going from the lowest ($/2$) to the highest ($\times 2$) rate decreases the surface $Z>35$ [X/Fe] on average by 0.3-0.4 dex. However,  lowering the rate by a factor of two does not affect the neutron production timescale enough to delay the i-process with respect to the splitting of the convective thermal pulse or to impact the surface enrichment \citep{Choplin21}.

\section{Sensitivity to neutron capture rates}
\label{sect_ng}

As far as reactions on heavier nuclei are concerned, most of the low-energy cross section calculations for practical applications are based on the
statistical model of Hauser-Feshbach (HF). Such a
model makes the fundamental assumption that the capture process takes place
with the intermediary formation of a compound nucleus (CN) in thermodynamic
equilibrium. The energy of the incident particle is then shared more or
less uniformly by all the nucleons before releasing the energy by
particle emission or $\gamma$-de-excitation. The formation of a CN
 is usually justified by assuming that the level density in the
compound nucleus at the projectile incident energy is large enough to
ensure an average statistical continuum superposition of available
resonances. The statistical model has shown to be able to
predict cross sections accurately. However, this model suffers from
uncertainties mostly stemming from the predicted nuclear
ingredients describing the nuclear structure properties of the ground
and excited states, as well as the strong and electromagnetic interaction
properties. In the case of radiative neutron captures for nuclei close to the valley of $\beta$-stability,
the two most uncertain inputs are clearly the nuclear level densities, and the photon strength functions.
Additionally, the direct capture (DC) may contribute to the reaction mechanism
for nuclei for which the number of available states in the CN is small \citep{Xu14}.

Our fiducial i-process nucleosynthesis calculations are performed with radiative neutron capture
rates computed with the {\sf TALYS} code within the HF statistical model for the 793 nuclei with $14 \le Z \le 93$ included in the i-process network and for which no experimental data exist.
In such calculations, the nuclear structure properties are obtained from the HFB-31 mass table \citep{Goriely16a},
except for the atomic masses and assignment of the ground-state spin and parities which are directly taken from known (or recommended) data \citep{Wang17,Audi17a}, whenever available.
The photon strength function is taken from the D1M+QRPA+0lim model corresponding to the quasi-particle random phase approximation (QRPA) based on
the Gogny D1M  interaction and including a non-zero E1 strength and M1 upbend at zero photon energy \citep{Goriely19}. The nuclear level densities
are taken from the Hartree-Fock-Bogolyubov (HFB) plus combinatorial calculations \citep{Goriely08b} and the optical potential from the
Woods-Saxon-type potential of \cite{Koning03}.

In the following sections (Sects.~\ref{sect_psf}-\ref{sect_nld}), the two major ingredients affecting the HF reaction rates, namely the photon strength function and the nuclear level densities, are coherently modified on the basis of alternative models to our fiducial one. In Sect.~\ref{sect_dc}, the impact of the DC component to the total capture rate is discussed. In all these cases, the neutron capture rates are recomputed coherently with the {\sf TALYS} reaction code and the impact of the correlated uncertainties on the i-process nucleosynthesis is analysed.  For the neutron densities found during the proton ingestion in our 1~\Msun{}, [Fe/H]=-2.5 model star, that is $N_n \la 10^{15}$~cm$^{-3}$, most of the neutron-rich nuclei produced during the irradiation have known $\beta$-decay rates. The impact of the still unknown $\beta$-decay rates for some 110 nuclei with $Z \ge 66$ included in our network has also been studied; it was found to be rather insignificant and for this reason will not be highlighted in the present study.
In addition to those 110 nuclei with unknown $\beta$-decay rates, 106 nuclei are known to have $\beta$-decay rates that are relatively sensitive to the temperature and density \citep{Takahashi87,Goriely99}.  The sensitivity of the i-process nucleosynthesis to those 106 $\beta$-decay rates has not been tested because those mostly affect nuclei close to the valley of $\beta$-stability on timescales longer than those found during the neutron irradiation. A dedicated investigation for the temperature- and density-dependence of the $\beta$-decay rates for additional neutron-rich nuclei potentially produced during the i-process is still missing.
Finally, about 159 nuclei in our reaction network have unmeasured masses. These concern $Z\ge 58$ neutron-rich nuclei but we have checked that those nuclei are not produced in any significant quantity, and therefore have a negligible impact on the i-process nucleosynthesis.

\subsection{Photon strength functions}
\label{sect_psf}
\begin{figure}
\includegraphics[width=\columnwidth]{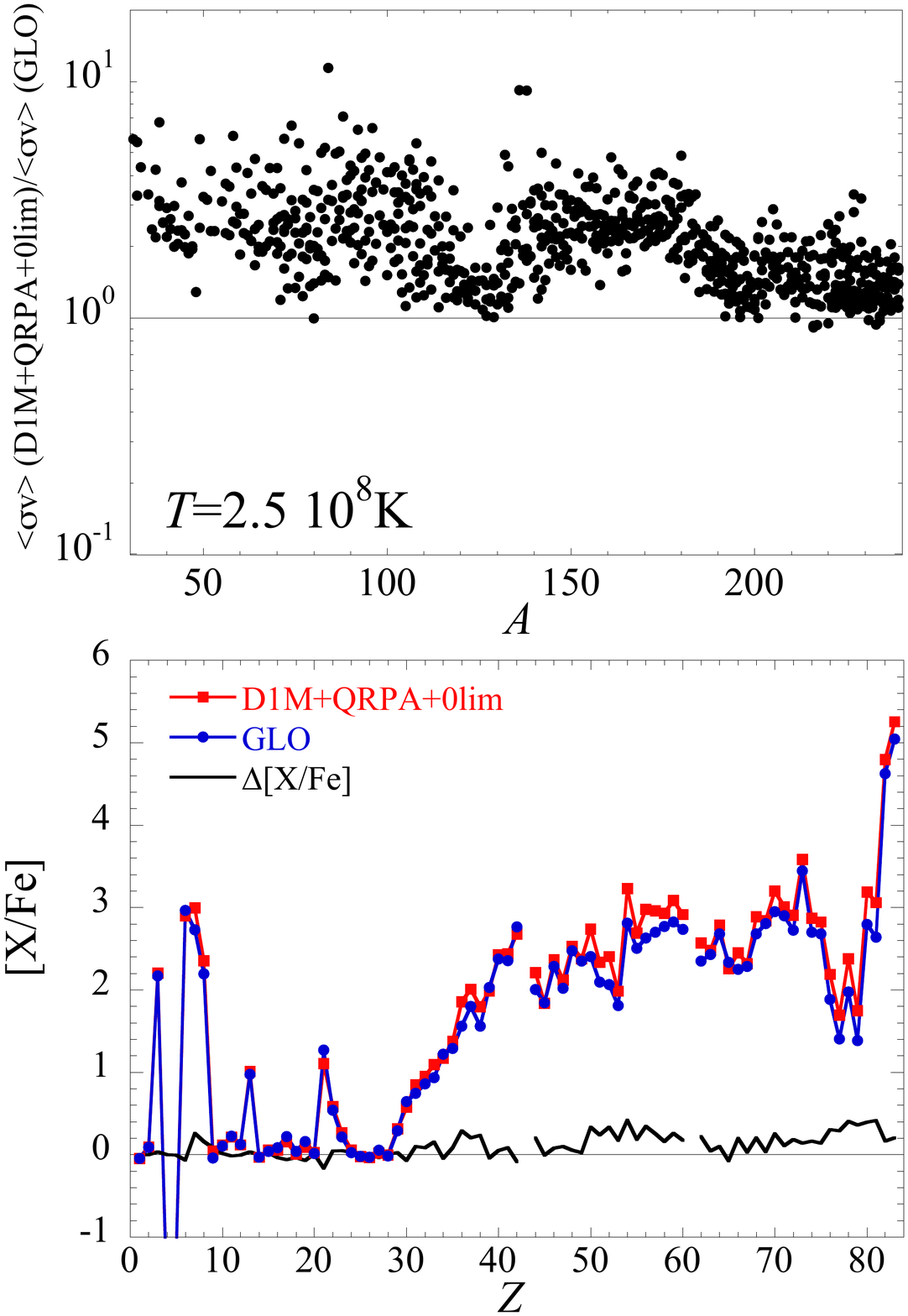}
\caption{{\it Upper panel}: Ratio between the neutron capture rates at $T = 2.5~10^8$~K obtained with the D1M+QRPA+0lim \citep{Goriely18a} and those obtained with the GLO \citep{Kopecky90,Capote09} photon strength functions, for all the 793 nuclei with $14 \le Z \le 93$ included in the i-process network and for which no experimental data exist.
{\it Lower panel}: Surface elemental overabundances resulting from the i-process in the 1~\Msun{}, [Fe/H]=-2.5 model star when adopting {\sf TALYS} radiative neutron capture rates obtained either with the D1M+QRPA+0lim or GLO photon strength functions. The black line corresponds to the difference in the overabundance between D1M+QRPA+0lim and GLO predictions.
 }
\label{fig_psf}
\end{figure}

The total photon transmission coefficient from a CN excited state is one of
the key ingredients for statistical cross section evaluation. The photon transmission
coefficient is  frequently described in the framework of the phenomenological
generalised Lorentzian (GLO) model of the giant dipole resonance \citep{Kopecky90,Capote09}.
Recent calculations within the QRPA approach based on a realistic Skyrme
or Gogny interaction have been shown to reproduce
the dipole $E1$ and $M1$ photon strength functions derived experimentally 
with the same level of accuracy \citep{Goriely19}.
Both models may, however, give different predictions for experimentally unknown nuclei, as shown in Fig.~\ref{fig_psf} (upper panel). More specifically, the D1M+QRPA+0lim model \citep{Goriely18a} is seen to give $(n,\gamma$) rates of up to a factor of ten higher than the GLO model \citep{Kopecky90,Capote09} which is due to the presence of low-lying strength in the QRPA approach but also to the inclusion of an $M1$ upbend component at low energies introduced to describe the de-excitation nature of the radiative capture. Faster rates tend to increase the overall enrichment, as shown in the lower panel of Fig.~\ref{fig_psf}. A global increase in the surface enrichment by 0.2-0.4 dex can be observed, predominantly for odd-$Z$ elements. This result is found to be independent of the adopted nuclear level density model.
Additional photon strength models, namely the Skyrme-HFB plus QRPA of \citet{Goriely04} and the simplified model Lorentzian (SMLO) of  \citet{Goriely18b} have also been tested. The latter provides rather similar results to our standard D1M+QRPA+0lim model. The Skyrme-HFB plus QRPA also gives similar predictions within typically 0.1--0.2 dex, except for Sc where a lower abundance by 0.4 dex is found with respect to D1M+QRPA+0lim.
The $M1$ upbend, as included in the D1M+QRPA+0lim model and described in \citet{Goriely18a}, is found to affect the $(n,\gamma$) rates of our 793 unmeasured nuclei by no more than 20\% to 40\% which barely impacts the i-process nucleosynthesis.

\subsection{Nuclear level densities}
\label{sect_nld}

\begin{figure}
\includegraphics[width=\columnwidth]{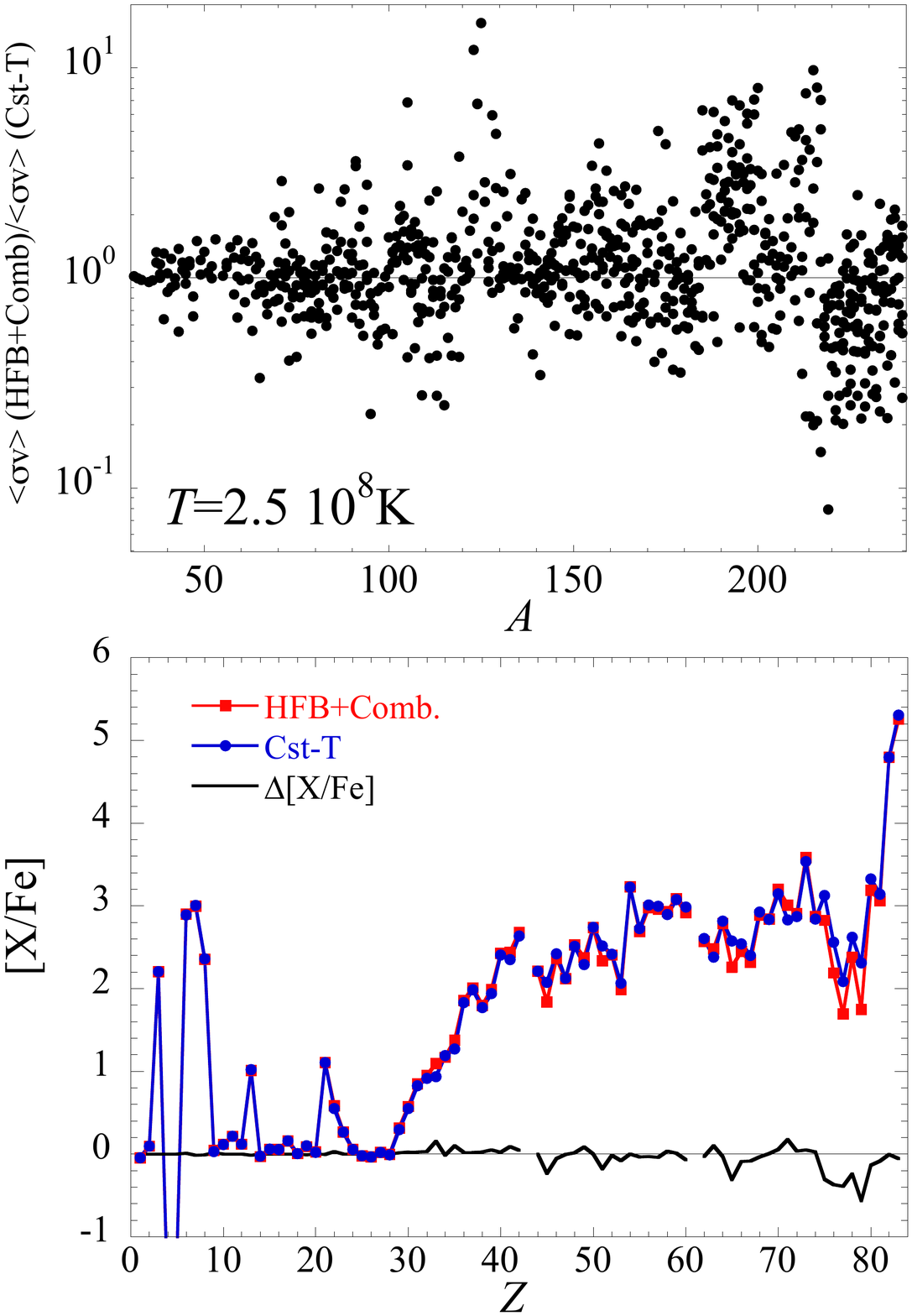}
\caption{Same as Fig.~\ref{fig_psf} when {\sf TALYS} radiative neutron captures are calculated with the nuclear level density models from either the HFB plus combinatorial (HFB+Comb) \citep{Goriely08b} or the constant-temperature (Cst-T) plus Fermi Gas model \citep{Koning08}. }
\label{fig_nld}
\end{figure}

Nuclear level densities are known to play an essential role in reaction theory.
Until recently, only classical analytical models of level densities were used for practical
applications. In particular, the  back-shifted Fermi gas model (BSFG) --or
some variant of it-- remains the most popular approach to estimating the spin-dependent
level densities, particularly in view of its ability to provide a simple analytical formula \citep{Capote09,Koning08}.
However, none of the important shell, pairing, and deformation effects are properly
accounted for in any analytical description and therefore large uncertainties are
expected, especially when extrapolating to very low energies (a few MeV) and/or to nuclei far from the valley of $\beta$-stability.
Several approximations used to obtain the level density expressions in an analytical form
can be avoided by quantitatively taking into account the discrete structure of the
single-particle spectra associated with realistic average potentials. This approach has
the advantage that shell, pairing, and deformation effects on all
the thermodynamic quantities are treated  in a natural way.
Large-scale calculations of nuclear level densities for nearly 8500 nuclei were performed in the framework of the combinatorial method  \citep{Goriely08b,Hilaire12} and  this latter has proven its predictive power. One of the main advantages of the combinatorial approach is to provide nuclear level densities not only  as a function of the excitation energy, but also of the spin and parity without any statistical assumption. The reaction rates obtained with the combinatorial method are compared with those obtained with the constant-temperature plus Fermi Gas model in Fig.~\ref{fig_nld} for the 793 nuclei of interest. Deviations by up to a factor up to ten can be observed. The overall impact on the i-process overabundances reaches up to 0.4 dex between both sets of rates. These are dominantly located in the Os-Au region.

\subsection{Direct capture}
\label{sect_dc}

When the number of available states in the CN is relatively small, the capture reaction may be dominated by direct electromagnetic transitions to a bound final state rather than through a compound nucleus intermediary. This DC proceeds via the excitation of only a few degrees of freedom on much shorter timescales reflecting the time taken by the projectile to travel across the target. This mechanism can be satisfactorily described with the perturbative approach known as the potential model \citep{Lynn68,Xu14}. It is now well accepted that the DC is important, and often dominant at the very low energies of astrophysical interest for light or neutron-rich systems for which few resonant states are available.

The three reaction mechanisms, that is, the CN, pre-equilibrium (PE), and DC,  have been studied systematically and comprehensively within a unique and consistent framework obtained with the {\sf TALYS} code \citep{Xu14}.  Importantly, the same nuclear inputs are used consistently to determine the three contributions. In particular, adopting the same nucleon--nucleus optical potential ensures that the three components are calculated on the same footing and represent partial fluxes of the same total reaction cross section. At  temperatures of relevance to the i-process ($T\simeq 2.5~10^8$~K), the PE contribution does not affect the total radiative capture rate \citep{Goriely08a}. The DC component is calculated here using an energy-independent spectroscopic factor inspired from shell model calculations \citep{Sieja21}.

\begin{figure}
\includegraphics[scale=0.25]{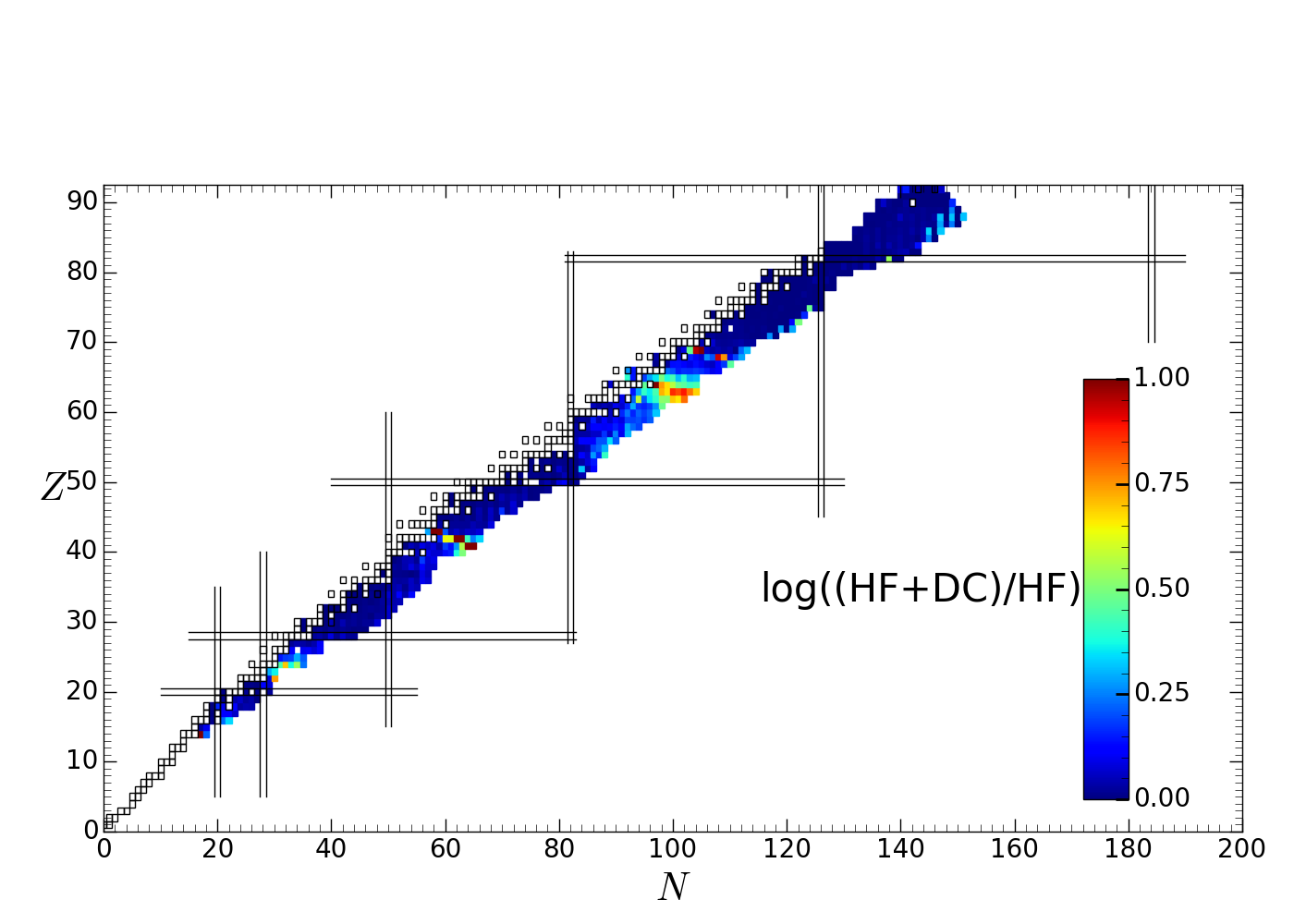}
\includegraphics[scale=0.32]{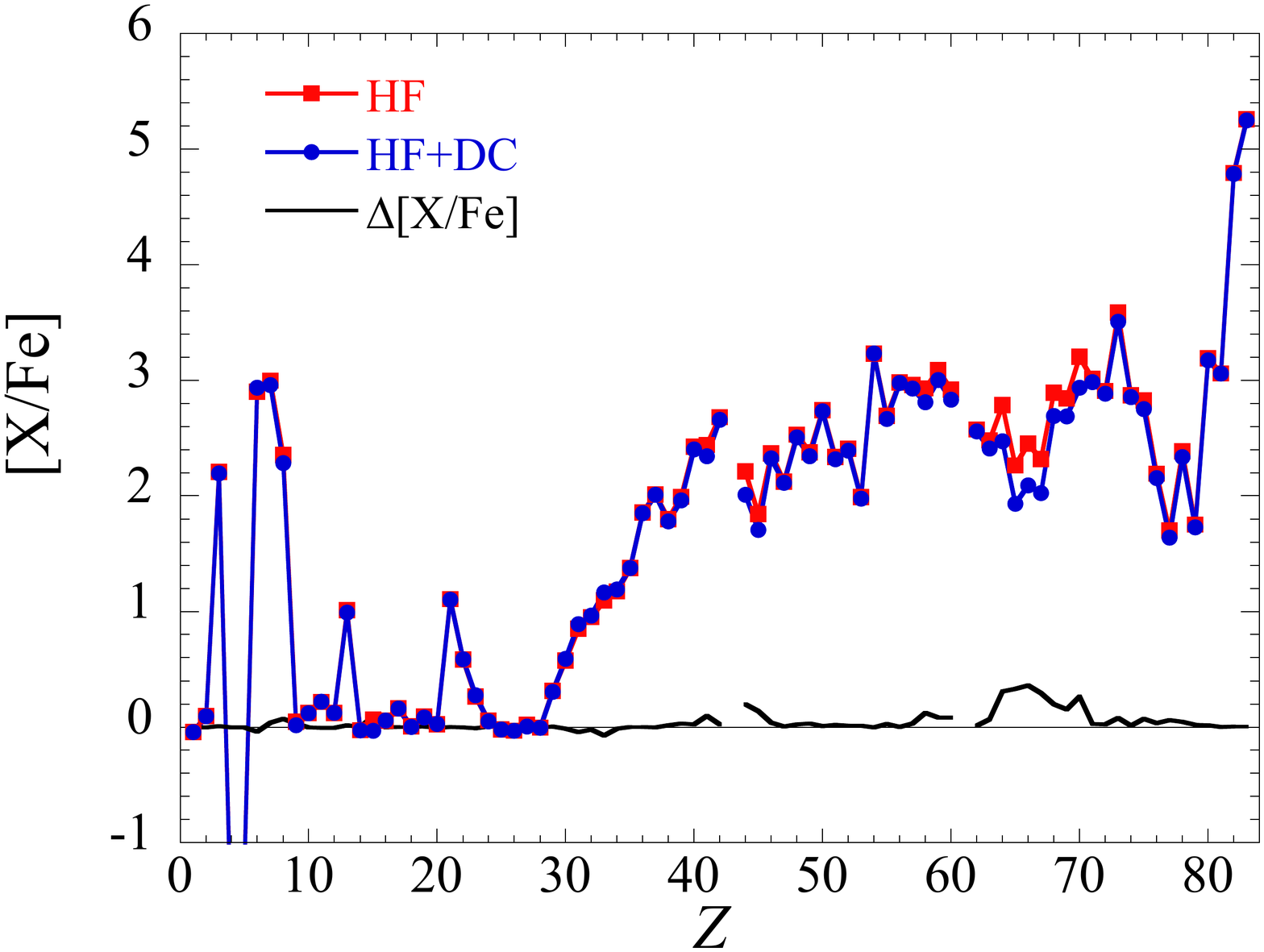}
\caption{{\it Upper panel}: Representation in the ($N,Z$) plane of the ratio between the total (HF + DC) and the HF neutron
capture rates at $T = 2.5~10^8$~K for all the 793 nuclei with $14 \le Z \le 93$ included in the i-process network and
for which no experimental data exist. Black open squares correspond to stable nuclei.
{\it Lower panel}: Surface elemental overabundances resulting from the i-process in the 1~\Msun{}, [Fe/H]=-2.5 model star when adopting {\sf TALYS} radiative neutron-capture rates excluding (HF) or including (HF+DC) the DC component to the reaction mechanism. The black line corresponds to the difference in the overabundance between HF+DC and DC predictions.
}
\label{fig_ng_dc}
\end{figure}

As shown in Fig.~\ref{fig_ng_dc}, the DC increases the radiative neutron capture rate by a factor of up to ten for some of the nuclei included in the i-process network. Figure~\ref{fig_ng_dc} shows that $A\simeq 31$, $A\simeq 100$ and $150 \la A \la 170$ regions in between closed neutron shells are dominated by the DC mechanism even close to the valley of $\beta$-stability. For this reason, the inclusion of the DC component particularly affects the elements around Rh ($Z=45$) and Tb ($Z=65$).


\section{Discussion}
\label{sect_disc}

\begin{figure}
\includegraphics[width=\columnwidth]{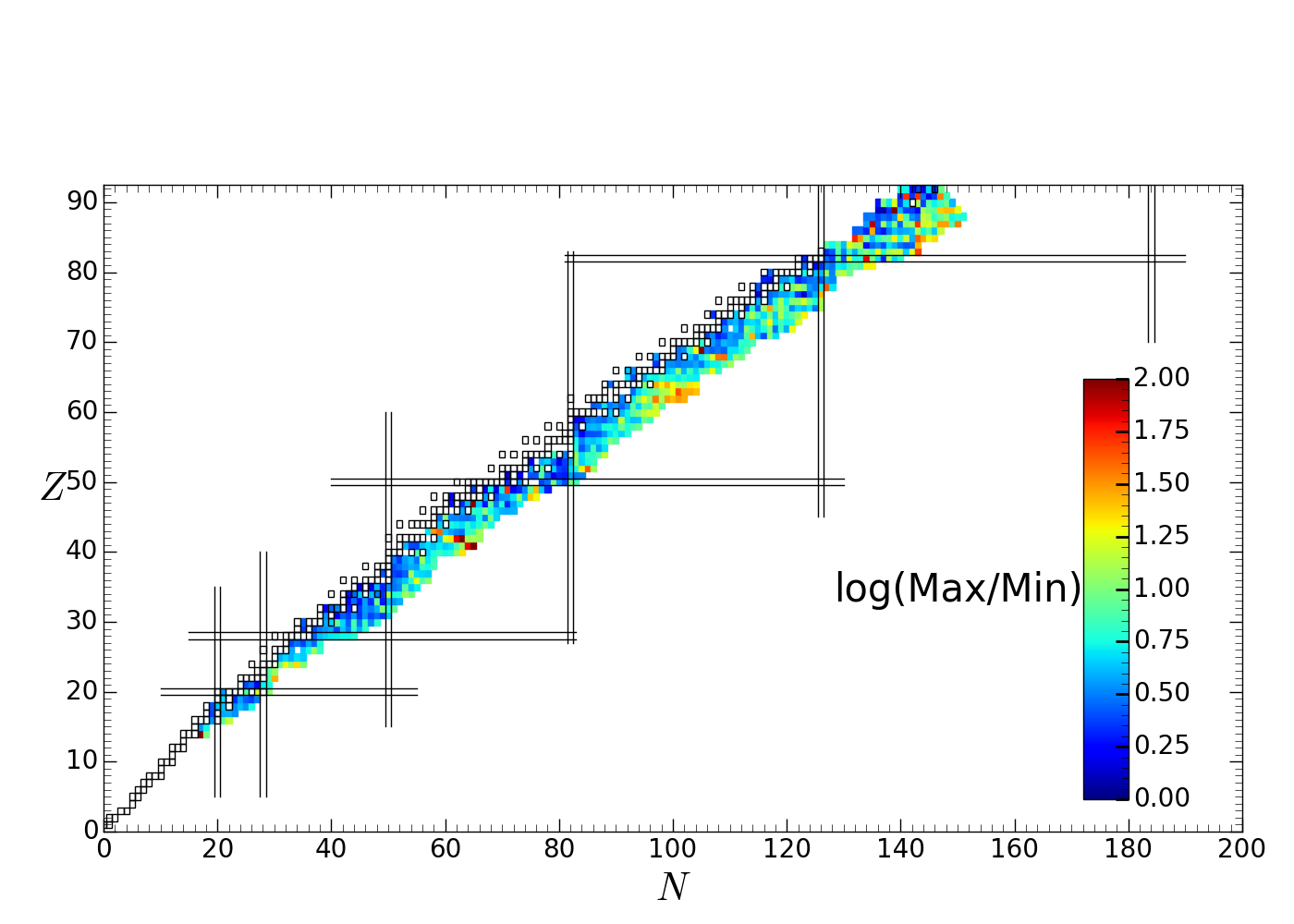}
\caption{Representation in the ($N,Z$) plane of the ratio between the maximum and minimum estimates of the neutron
capture rates at $T = 2.5~10^8$~K between the ten {\sf TALYS} sets for all the 793 nuclei with $14 \le Z \le 93$ included in the i-process network and
for which no experimental data exist. The black open squares correspond to stable nuclei.
}
\label{fig_ng_tot}
\end{figure}

As discussed in Sect.~\ref{sect_ng}, uncertainties in photon strength functions and nuclear level densities may affect the predictions of experimentally unknown rates. We considered ten different sets of {\sf TALYS} predictions and estimated the corresponding surface overabundances resulting from the i-process in the 1~\Msun{}, [Fe/H]=-2.5 model star. The amplitude of the changes in the radiative neutron capture rates for each of the 793 nuclei is illustrated in Fig.~\ref{fig_ng_tot} where variations by factors of up to about 100 can be observed. It is important to emphasise that all the sets considered here reproduce the 240 experimental Maxwellian-averaged neutron-capture cross sections \citep{Dillman06} for nuclei with $20 \le Z \le 83$ with a root mean square deviation of the ratio of no more than 1.7.  This means that the accuracy of all the sets is rather high and there is no reason to reject any of these ten sets.

\begin{figure}
\includegraphics[width=\columnwidth]{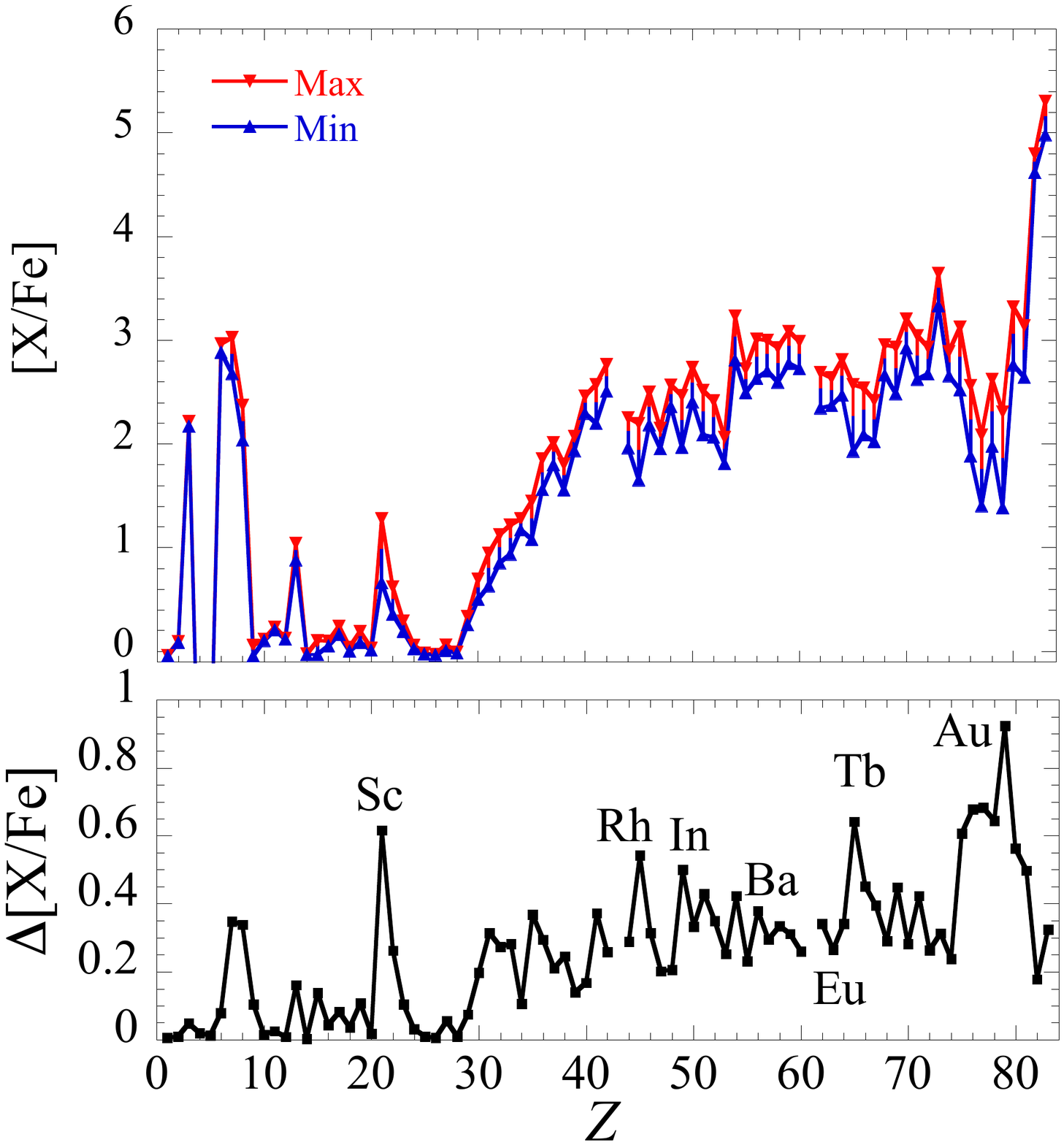}
\caption{Upper and lower estimates of the surface elemental overabundances resulting from the i-process in the 1~\Msun{}, [Fe/H]=-2.5 model star when adopting ten different sets of {\sf TALYS} reaction rates.}
\label{fig_ipro_tot}
\end{figure}

\begin{figure}
\includegraphics[width=\columnwidth]{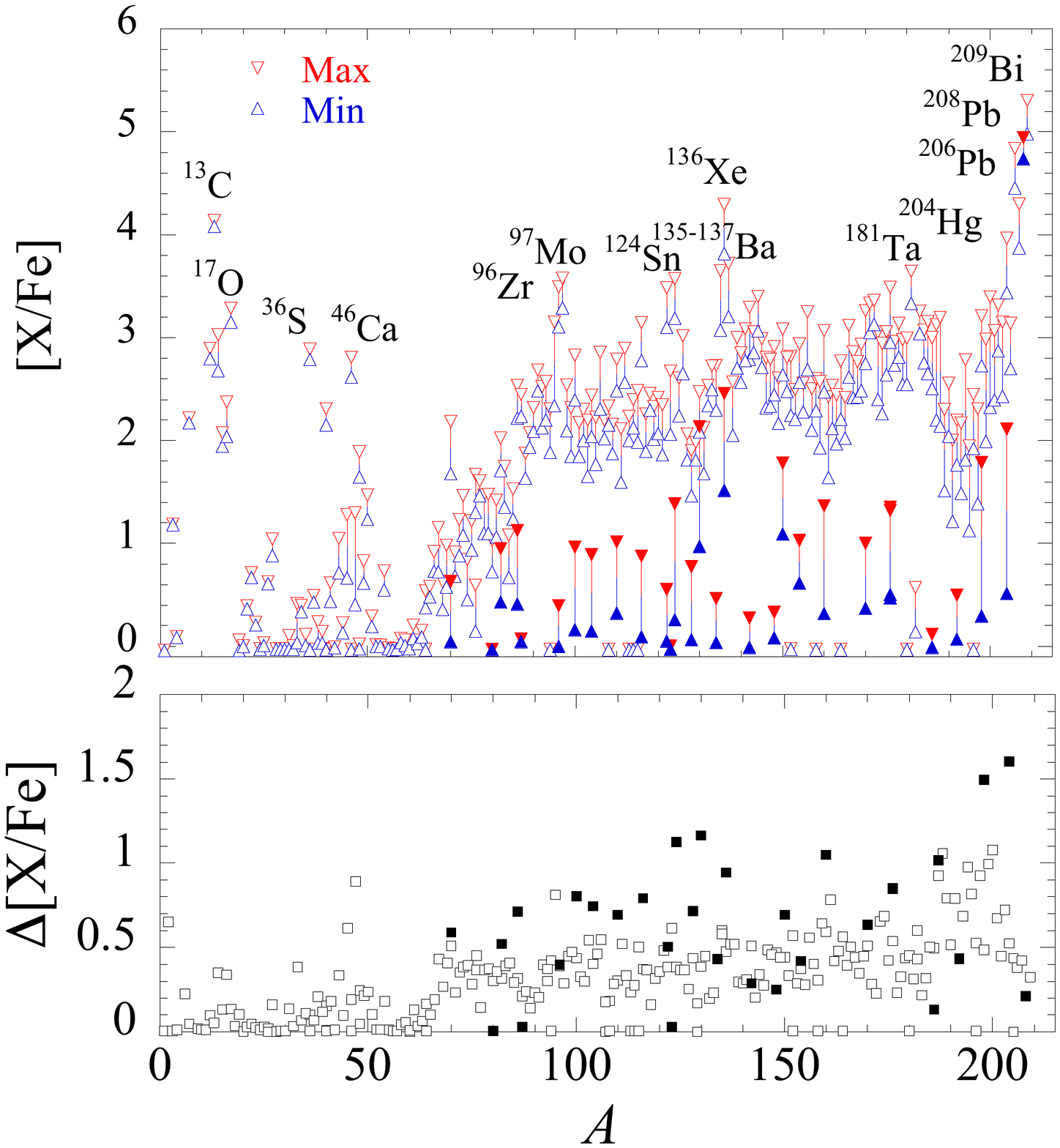}
\caption{Same as Fig.~\ref{fig_ipro_tot} but for the surface isotopic overabundances. The s-only nuclei are shown by solid symbols; except for Pb, their abundances are significantly lower than those of the sr-nuclei, which are indicated by open symbols.}
\label{fig_ipro_tot_ZA}
\end{figure}

The resulting upper and lower limits to the surface  elemental distribution are  shown in Fig.~\ref{fig_ipro_tot}, while those limits to the surface  isotopic distribution are shown in Fig.~\ref{fig_ipro_tot_ZA}. A word of caution should be given here. While the uncertainties have been consistently propagated to the i-process nucleosynthesis, the upper and lower limits shown in Figs.~\ref{fig_ipro_tot}-\ref{fig_ipro_tot_ZA} do not correspond to two specific sets and the correlation between the model uncertainties is hidden somewhere in between those two limits. Figures~\ref{fig_ipro_tot}-\ref{fig_ipro_tot_ZA} illustrate, however, the scale of the impact that can be expected from such model uncertainties. The highest overabundances are mainly obtained with the HF model together with the D1M+QRPA+0lim photon strength function while the lowest predictions are found when including the DC component.

Interestingly, the $56 \la Z \la 63$ region of the spectroscopically relevant heavy-s elements of Ba-La-Ce-Pr  as well as the r-dominated element Eu are seen to remain weakly affected by nuclear uncertainties. Nuclear models in this region close to the $N=82$ neutron shell may give rise to different reaction rates (as seen in Fig.~\ref{fig_psf}), but the correlation between the uncertainties never significantly alters  the nucleosynthesis. A similar conclusion is found for the light $Z<30$ elements, except for Sc.

\begin{figure}
\includegraphics[width=\columnwidth]{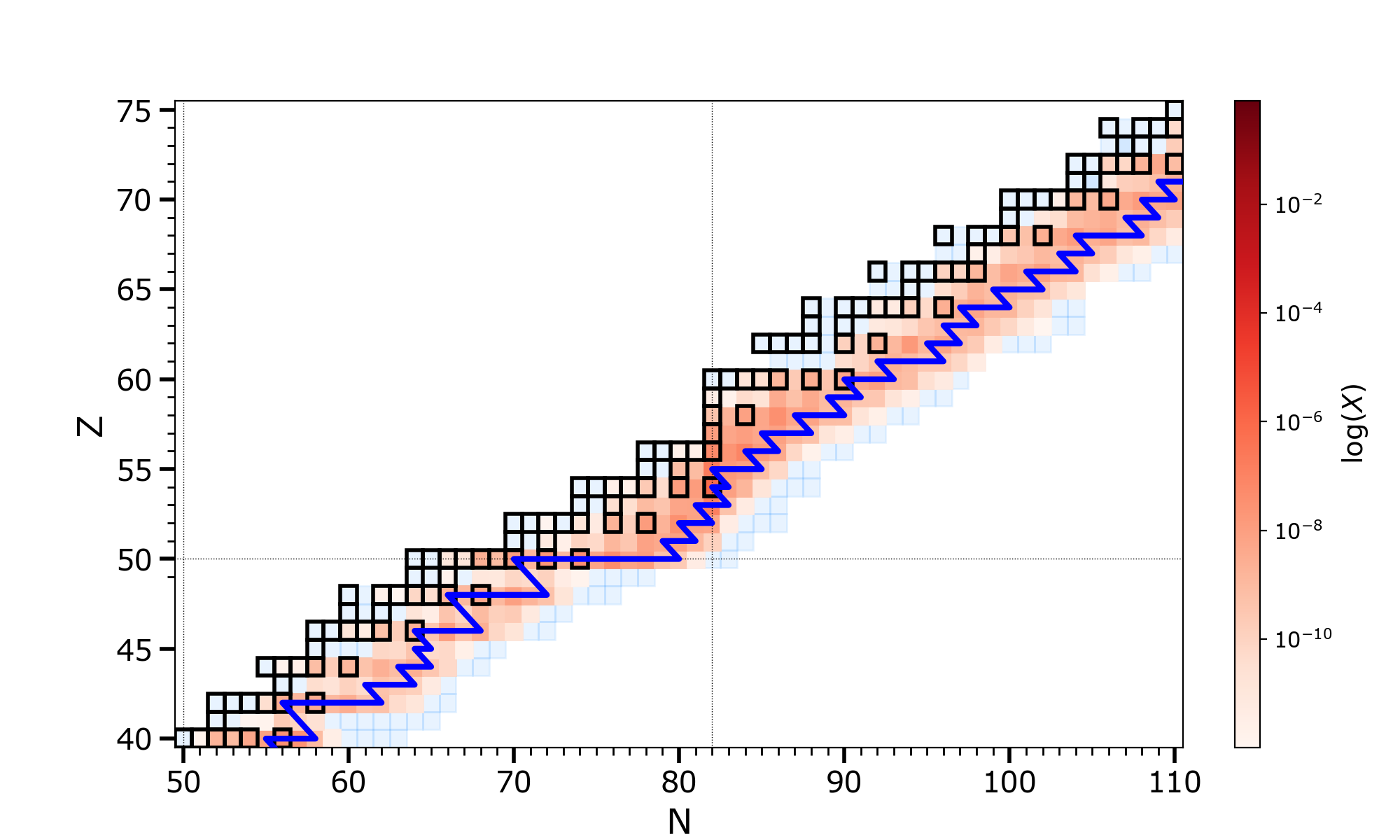}
\caption{Illustration of the main nuclear i-process flow (blue line) in the layer of 1~\Msun{}, [Fe/H]=-2.5 model star with the highest neutron density of $N_n \simeq 5 \times 10^{14}$~cm$^{-3}$. The blue squares correspond to the nuclei included in the network, the black ones to the stable nuclei. The abundances are depicted by the red colour scale in mass fraction. The $N=82$ isotones and $Z=50$ isotopes are highlighted by the vertical and horizontal dotted lines,
respectively.}
\label{fig_flow}
\end{figure}

One of the regions with the highest impact of nuclear uncertainties is located around $Z=65$. This region is particularly affected by the increase in neutron capture through the DC contribution which decreases the production of the $64 \la Z \la 67$ elements, as shown in the lower panel of Fig.~\ref{fig_ng_dc}. A similar effect is found in the $Z=45$ region. The estimate of the DC contribution requires a detailed account of the excited spectrum, including the spectroscopic factors. Experimental constraints from ($d,p$) reactions or dedicated shell-model calculations in the precursor regions of $^{102-104}$Mo, $^{160}$Sm, or $^{160}$Eu (see Fig.~\ref{fig_flow}) could help to reduce the uncertainties on the DC contribution. Uncertainties in the corresponding photon strength function also add to the large impact on the overabundance factors.

As far as level densities are concerned, they tend to affect abundance predictions mainly in the $Z=74-79$ regions, as illustrated in Fig.~\ref{fig_nld}. Experimentally derived nuclear level densities at low energies have been shown to be compatible with a constant-temperature formula \cite[e.g.][]{Giacoppo14,Tornyi14}, though it is possible to show that the HFB plus combinatorial approach can also explain all the Oslo measurements with a relatively similar accuracy \citep{Goriely08b}. However, differences in both approaches in the neutron-rich region of $^{191}$Re up to $^{197}$Os are responsible for the main deviations found in the nucleosynthesis calculations. Stronger constraints on nuclear level densities in this region would improve the i-process predictions.

As far as the isotopic abundance distribution is concerned (Fig.~\ref{fig_ipro_tot_ZA}), deviations of up to 1.5 dex can be obtained. The enrichment of s-only nuclei, that is, nuclei expected to be exclusively produced by the s-process, are found to be significantly affected by nuclear uncertainties, but, except for $^{208}$Pb, their overall overabundances are significantly lower than those of sr-nuclei. In particular, the odd-$A$ s-only nuclei are not overproduced, and $^{187}$Os is found to be significantly destroyed by the i-process.

\section{Conclusions}
\label{sect_conc}
Our first paper \citep{Choplin21} confirmed that the ingestion of protons in the convective helium-burning region of low-mass low-metallicity stars can explain the surface abundance distribution observed in CEMP r/s stars  relatively well. In this first study, the i-process sensitivity to the stellar evolution numerics was analysed. It was shown in particular that proper temporal and spatial resolutions are needed for a reliable prediction of the i-process nucleosynthesis. However,  the i-process reaction network calculations also involve hundreds of nuclei for which nuclear reaction or decay rates remain affected by large uncertainties, especially those not available experimentally, and this could impact nucleosynthesis predictions. The aim of the present study was to carry out a detailed analysis of the impact of  nuclear uncertainties on the surface enrichment following the i-process in a low-mass low-metallicity AGB star. To do so, we computed the evolution of a 1 \Msun\ model at [Fe/H]~$=-2.5$ during a proton ingestion event using a nuclear network of 1160 species coupled to the transport processes and propagated nuclear uncertainties consistently.

Uncertainties associated with unknown $\beta$-decay rates or atomic masses are found to have an insignificant impact on the i-process nucleosynthesis, at least for nuclear flow limited to neutron densities of the order of $10^{15}$~cm$^{-3}$. A modification of the  \iso{13}C(\an)\iso{16}O reaction rate is also found to have a relatively low impact on the overall surface enrichment, in particular concerning a possible delay of the nucleosynthesis with respect to the split of the convective zone.

We thoroughly investigated the nuclear physics uncertainties associated with theoretically derived radiative neutron capture cross sections for some 793 nuclei in the reaction network. We consistently calculated different sets of radiative neutron capture cross sections with the  {\sf TALYS} reaction code by adopting different models of the photon strength functions and nuclear level densities entering the calculation of resonant cross section and accounting for the possible contribution from the direct capture component.  The corresponding global systematic uncertainties were propagated to the calculation of reaction rates, and, keeping the correlation imposed by the underlying model,  were consistently applied to the nucleosynthesis simulations to estimate their impact on the i-process yields.
It is found that nuclear uncertainties affect surface elemental abundances by typically 0.4~dex
on average. The inclusion of the direct capture contribution impacts the rates in the $A\simeq 45$, 100, 160, and 200 regions, and the i-process production of the $Z\simeq 45$ and 65-70 elements. Uncertainties in the corresponding photon strength function also contribute to the impact on the abundance uncertainties. Nuclear level densities tend to affect abundance predictions mainly in the $Z=74-79$ region. In contrast, the  $56 \la Z \la 59$ region of the spectroscopically relevant heavy-s elements of Ba-La-Ce-Pr  as well as the r-dominated element Eu are  only slightly affected by nuclear uncertainties. The enrichment of s-only nuclei are found to be significantly affected by nuclear uncertainties, but, with the exception of $^{208}$Pb, their overall isotopic overabundances remain negligible compared to their corresponding sr-nuclei isotopes.

Experimental constraints from ($d,p$) reactions or dedicated shell-model calculations in the precursor regions of $^{102-104}$Mo, $^{158}$Sm, or $^{160}$Eu (see Fig.~\ref{fig_flow}) could help to reduce the uncertainties on the DC contribution. Similarly, Oslo-type measurements to constrain the photon strength functions and nuclear level densities, in particular in the slightly neutron-rich $A\simeq 100$, 160, and 197 regions, could improve the predictive power of the i-process calculations. Similar sensitivity studies of the nuclear uncertainties with a consistent propagation to the nucleosynthesis calculation still need to be performed in other plausible i-process sites, including AGB model stars of different masses and metallicities that would be subject to similar large neutron irradiations.

\section*{Acknowledgments}
This work was supported by the Fonds de la Recherche Scientifique-FNRS under Grant No IISN 4.4502.19.
L.S. and S.G. are senior FRS-F.N.R.S. research associates.

\bibliographystyle{aa}
\bibliography{astro}

\end{document}